\documentclass[12pt,preprint]{aastex}

\def\lya{\ifmmode {\rm Ly}\alpha~ \else Ly$\alpha$~\fi}
\def\lyb{\ifmmode {\rm Ly}\beta~ \else Ly$\beta$~\fi}
\def\lyg{\ifmmode {\rm Ly}\gamma~ \else Ly$\gamma$~\fi}
\def\civ{\ifmmode {\rm C}\,{\sc iv}~ \else C\,{\sc iv}~\fi}
\def\civn{\ifmmode {\rm C}\,{\sc iv}~ \else C\,{\sc iv}\fi}
\def\cvn{\ifmmode {\rm C}\,{\sc v}~ \else C\,{\sc v}\fi}
\def\cvin{\ifmmode {\rm C}\,{\sc vi}~ \else C\,{\sc vi}\fi}
\def\nvn{\ifmmode {\rm N}\,{\sc v}~ \else N\,{\sc v}\fi}
\def\nvin{\ifmmode {\rm N}\,{\sc vi}~ \else N\,{\sc vi}\fi}
\def\nviin{\ifmmode {\rm N}\,{\sc vii}~ \else N\,{\sc vii}\fi}

\def\ov{{{\rm O}\,\hbox{{\sc v}}~}}

\def\ovii{{{\rm O}\,\hbox{{\sc vii}}~}}
\def\oviii{{{\rm O}\,\hbox{{\sc viii}}~}}

\def\neix{\ifmmode {\rm Ne}\,{\sc ix}~ \else Ne\,{\sc ix}~\fi}
\def\nex{\ifmmode {\rm Ne}\,{\sc x}~ \else Ne\,{\sc x}~\fi}
\def\hi{\ifmmode {\rm H}\,{\sc i}~ \else H\,{\sc i}~\fi}

\def\chandra {{\it Chandra}~}

\begin{document}
\title{
The weak absorbing outflow in AGN Mrk 279: evidence of super-solar metal
abundances. }
\author{Dale L. Fields\altaffilmark{1,2}, Smita Mathur\altaffilmark{1}, 
Yair Krongold\altaffilmark{3,4},  Rik Williams\altaffilmark{1,5}, \& 
Fabrizio Nicastro\altaffilmark{3} }
\altaffiltext{1}
{Department of Astronomy, the Ohio State University, 140 West 18th
Avenue, Columbus, OH 43210, USA}
\altaffiltext{2}
{Current address: Pierce College, 6201 Winnetka Ave., Woodland Hills, CA 91371}
\altaffiltext{3}
{SAO,
60 Garden Street, 02138, Cambridge, MA, USA}
\altaffiltext{4}
{Instituto de Astronom\'ia Universidad Aut\'onomica de M\'exico,
Apartado Postal 70 - 264, Ciudad Universitaria, M\'exico, D.F., CP
04510, M\'exico}
\altaffiltext{5}
{Current address: Leiden Observatory, Lieden University, P.O.Box 9513,
2300 RA Leiden, The Netherlands}

\email{smita@astronomy.ohio-state.edu, FieldsDL@piercecollege.edu}

\begin{abstract}
We present analysis and photoionization modeling of the \chandra high
resolution spectrum of Mrk~279. There is clear evidence of an absorbing
outflow which is best fit by a two component model, one with a low
ionization parameter and one with a higher ionization parameter. The
column density of the X-ray warm absorber, about $\log N_H=20$, is the
smallest known of all AGNs in which X-ray absorbing outflows are
observed. We find that the X-ray and UV/FUV absorbers are part of the
same overall outflow. There is some evidence of super-solar carbon,
nitrogen, oxygen and iron in the \chandra spectrum of Mrk 279. While
this is not a robust result in itself, \chandra data in combination with
the UV data and the pressure equilibrium between two phases of the
outflow, support the scenario of super-solar abundances. This the first
case where super-solar abundances are reported in the nucleus of a
normal Seyfert galaxy. The data suggest that the outflow originates from
a compact region around the nuclear black hole and that it carries
insignificant amount of mass and energy.
\end{abstract}


\section{Introduction} \label{sec:intro}

Feedback from quasars has been a key word in recent years as a way of
solving many astrophysical problems ranging from cluster cooling flows
to structures of galaxies. Quasars can provide energy to the surrounding
medium in two ways, either through their radiative luminosity or through
outflows. Strong continuum emission from quasars plays a significant
role in providing the background radiation that keeps the intergalactic
medium ionized at redshifts around three. Outflows from quasars
themselves come in two different forms, seen as radio jets and as
absorbing outflows. The feedback from jets can be sufficient to keep the
cooling flows in clusters from cooling further to much lower temperatures
(e.g. McNamara et al. 2001) or to regulate black hole growth
\citep{rafferty} in cluster dominant galaxies. Only about 10\% of all
quasars, however, are radio-loud and may have radio jets, so jet-related
feedback cannot be common. In a fraction of mostly radio-quiet quasars,
absorption lines of highly ionized metals are observed blue-shifted with
respect to the corresponding emission lines. These ``absorbing
outflows'' then provide us with the most common form of possible quasar
feedback. The outflows from active galactic nuclei (AGNs) might also be
responsible for enriching the intergalactic medium with metals,
especially if they are strong and contain high metallicity
gas. Understanding the physical conditions in the absorbing components,
and their mass and energy outflow rates, thus becomes very important.

Absorbing outflows have been studied in great detail in lower redshift,
lower luminosity, Seyfert galaxies.  Absorbing outflows exist in about
50\% of AGNs, as seen through UV absorption lines \citep{crenshaw99} and
X-ray ``warm absorbers'' \citep{r97,george}. In several of these AGNs,
the UV absorber was found to be related to the X-ray absorber
\citep{m94, m5548, m97, m98, m01}.  \chandra observations have
revolutionized our understanding of absorbing outflows through
detections of a large number of resonance lines, in addition to
edges. One of the best spectra of AGN warm absorbers observed with
\chandra is of NGC\,3783 \citep{kaspi} which showed over 100 absorption
lines from hydrogen-like and helium-like ions of multiple elements
\citep{phase, net03}.  This rich system was remarkably well-described
with a simple two- or three-phase absorbing medium in pressure balance.
In this paper we present results from the analysis of the \chandra
spectrum of Mrk\,279, a nearby Seyfert 1 galaxy (z=0.03).

Mrk\,279 was observed on May 2002 with HST, FUSE and \chandra
simultaneously, the results of which are published in
\citet{scott279}. These authors focus mainly on the HST and FUSE data
because no \ovii or \oviii lines were detected in the X-ray spectrum due
to the low S/N in the \chandra data, primarily because Mrk 279 happened
to be in a low state at the time of the observation.  A second observing
campaign, again with all three observatories \chandra, HST and FUSE
followed in May 2003. The results from the HST and FUSE observations are
published in \citet{gabel} and in \citet{arav}.  \citet{kaastra} discuss
the density diagnostic power of \ov K-shell absorption lines and apply
it to the Mrk\,279 X-ray spectrum. The detection of \ov in the \chandra
spectrum of Mrk\,279 is not secure, but if the absorption is indeed due
to \ov line, then the distance of the absorber from the continuum source
is constrained to be about one light week to a few light months.
\citet{c05} present the \chandra LETG spectrum of Mrk\,279 and report on
the presence of a two component absorption system intrinsic to the
source. In their preliminary analysis, presented in a conference
proceedings, they find a low ionization component with total column
density $\log N_H\sim 1.6\times 10^{22}$ cm$^{-2}$ and a higher
ionization component with column density twice as high. These authors
claim that the two absorption systems belong to two different outflow
velocity systems with $-220^{+50}_{-90}$ and $-570^{+100}_{-70}$\,km
s$^{-1}$ for the low and high ionization systems respectively.

We have recently developed a code called PHASE (discussed below) with
the goal of understanding photoionized plasma such as found in the
circumnuclear regions of AGNs. PHASE has been highly successful in its
application to the \chandra spectrum of NGC\,3783 \citep{phase,k04},
NGC\,985 \citep{k05}, and the XMM spectra of NGC\,4051 \citep{k06}. In
NGC\,3783 we found an absorbing outflow with two ionization components
in pressure balance with each other.  We were able to accurately
determine the size of the warm absorber for the first time in NGC\,4051
by combining the detailed spectral model from the XMM-Newton RGS data to
the variability observed in the XMM-Newton EPIC data. With the accurate
PHASE modeling, we are in a position to rule out some models of AGN
outflow such as those with continuous range of ionization parameters and
those in which absorption arises in extended, kpc-scale structures.

Here we present our PHASE analysis of the \chandra spectrum of Mrk\,279 from
the 2003 observing campaign.  With the long wavelength range observed,
multiple ionization states of many elements are detected, which allow us
to constrain the physical state of the absorber. The connection between
the X-ray and the UV absorber can also be tested.  Our goal is hence
multi-fold: (1) we wish to constrain the physical conditions in the
Mrk\,279 outflow using \chandra data, (2) compare the results to the UV
absorption system reported in \citet{gabel}, (3) compare our results
with the  results of \citet{c05}, and (4) compare the
Mrk\,279 system with other AGN outflows.

\section{Observations and Data Reduction} \label{sec:data}

Mrk 279 was observed seven times by the \chandra X-ray observatory in
May 2003, with the low energy transmission grating (LETG) and the high
resolution camera spectroscopic array (HRC-S), for a total of 340 ks of
exposure time. The full details of the observations and data reduction
are presented in \citet{rik} where the $z=0$ absorption system in the
line of sight to Mrk\,279 is discussed.  The focus of this paper is
instead on the intrinsic absorption line system.  Spectral orders from
$-6$ to $+6$ were included in the combined instrumental response matrix
in order to model the continuum accurately.  We note here that the
resulting coadded LETG spectrum has a signal-to-noise ratio of S/N $\sim
6.5$ near 22\AA, and the S/N is good over the wavelength range of
10--50\AA. The typical LETG/HRC-S resolution is about 0.05\AA\ (FWHM),
or 600 km s$^{-1}$ at 25\AA. The 10--50\AA\ spectrum is shown in Figure 1
of \citet{rik}.

\section{Analysis} \label{sec:analysis}
\subsection{Spectral Fit}

We conduct our analysis using {\it Sherpa} \citep{sherpa} which is a
part of the CIAO (Chandra Interactive Analysis of Observations, 
\citet{ciao}) package.  The global fit to the spectrum is performed using
our recently developed code PHASE (PHotoionized Absorber Spectral
Engine) which self-consistently reproduces the X-ray absorption spectrum
of an intrinsic absorber. Detailed description of PHASE is given in 
\citet{phase}; we describe it briefly here.  PHASE calculates absorption
due to ionized plasma using an extensive atomic database; the ionization
balance is calculated using the CLOUDY code (version 90.04; Ferland
1997). As in CLOUDY, the incident continuum is from the AGN and a simple
plane parallel geometry for the absorber is assumed. PHASE calculates both
bound-bound and bound-free transitions, with $\sim 4000$ lines including
Fe L-shell transitions and the Fe M-shell UTA (unresolved transition
array).  At its simplest, the data can be fit with PHASE with only three
input parameters: the equivalent hydrogen column density ($\log N_H$),
the ionization parameter ($\log U$), and the wavelength centroid
(expressed as the redshift of the absorber $z$).  This model is also
capable of fitting the velocity widths and the individual abundances of
most pre-iron elements, though these are less well constrained than the
primary parameters.

The observed continuum of Mrk 279 is reasonably fit with a single
powerlaw and Galactic absorption of $1.78\times10^{20}$\,cm$^{-2}$ as
found in \citet{rik}.  We allow the value of the Galactic absorption and
the powerlaw to fit freely.  The Galactic $N_H$ value is effectively
fixed as the variation in it is small from solution to solution.  The
powerlaw must be allowed to fit freely as high metal column densities
can have significant wide-band absorption effects.  We have given the
continuum the additional freedom of a broken powerlaw but the result is
a fit with the break at 0.06\,keV, below the observed energy range,
indicating that the additional freedom is unnecessary.  In all our
models we shall report results with a continuum of a single powerlaw
plus Galactic absorption.  To this continuum model we add one (and later
a second) PHASE absorber. Even with just the continuum model, the fit is
good ($\chi^2=3020$ for $3198$ degrees of freedom). As we show below,
however, the fit improves significantly by adding an absorbing
component. The strong 21.6\,\AA\ line in the spectrum of Mrk 279 is from
\ovii\ K$\alpha$ at redshift zero as reported by \citet{rik}.
In what follows we will discuss only the warm absorber intrinsic to
Mrk\,279.

\subsection{Mrk\,279 LIP Absorber} \label{sec:lip}

The most prominent absorption lines in the spectrum of Mrk\,279 are
those of \ion{C}{5}, \ion{N}{6}, \ion{O}{5}, \ion{O}{6}, and \ion{O}{7}
(Figure 1) which all indicate the presence of a low-ionization absorber,
similar to the low ionization phase (LIP) component observed in the
spectra of other warm absorbers (e.g. NGC 3783, \citealt{phase},
\citealt{net03}).  We therefore first try to fit the absorption spectrum
with a single absorber.  The quality of the fit will determine whether
we must add additional components, and whether an additional component
will improve the fit significantly.  For our single-absorber fit, we use
the PHASE model, and only allow three parameters to vary: the ionization
parameter $\log U$, the hydrogen column density $N_H$, and the absorber
redshift $z$; the continuum parameters are also allowed to vary.  We fix
the velocity width of the absorber to $100$\,km\,s$^{-1}$ as in
\citet{phase} and the abundances are fixed to solar. Changing the
velocity dispersion to smaller values did not significantly change our
results.

The existence of an absorber is heavily favored with a $\Delta\chi^2\sim
132$.  This absorber has best fit parameters of $\log U=-0.36$, $\log
N_H=19.9$, and an outflow velocity relative to systemic of $135$\,km
s$^{-1}$.  This solution is robust to initial conditions; we will refer
to this absorber solution as our model 1 (Table 1) with a low-ionization
phase (LIP) component.


\subsection{Mrk\,279 HIP Absorber} \label{sec:hip}

The best fit LIP single-absorber solution systematically under-predicts
several lines, the most prominent of which is \ion{O}{8} K\,$\alpha$
(see figure 2), and also under-predicts the amount of absorption blue-ward
of the Fe-UTA.  This observation, again, is similar to that in
NGC\,3783, and suggests the presence of a higher-ionized phase (HIP)
absorber component.  While the LIP solution can be modified to fit the
Hydrogen-like lines of carbon, nitrogen and oxygen by increasing the
column densities of these elements, doing so severely over-predicts the
amount of absorption due to the Helium-like ions.  For this reason we
add a second absorber to our model.  In a similar manner to the process
of fitting the LIP alone, we fit the LIP$+$HIP model with three new
degrees of freedom ($\log U_{HIP}$, $\log N_{H,HIP}$, and outflow
velocity $z_{HIP}$).

The addition of this component (with three new degrees of freedom; $\log
U$, $\log N_H$, and outflow velocity) has a $\Delta\chi^2$ of 43
indicating it is significant (F-test confidence of 99.999\%)(Figure
1). The solution has $\log U_{LIP}=-0.95$, $\log N_{H,LIP}=20.19$, and
$\log U_{HIP}=+0.77$, $\log N_{H,HIP}=20.16$. Typical $1\sigma$ errors
on $\log N_H$ are $\pm 0.1$ and on $\log U$ are about $^{+0.3}_{-0.05}$
for the LIP and $\pm0.2$ for the HIP. We refer to this solution as our
model 2 (Table 1). The large positive error in the ionization parameter
of the LIP is due to the lack of accurate low temperature dielectronic
recombination rate coefficients for Fe VII-XIII (see Krongold et
al. 2005 for a detailed discussion). These charge states are responsible
for absorption of the so-called unresolved transition array
(UTA). Even though the UTA is weak in the Mrk 279 data, it can still
produce a slight underestimation of U for this component. However, such
underestimation cannot be by more than a factor of 2, as shown by Netzer
(2004) and Kraemer et al. (2004).

Next we test the LIP$+$HIP absorption solution for its
sensitivity to the abundance of each element.  We find that the LIP is
most sensitive to the absorption lines of carbon, nitrogen, oxygen, and
the UTA of iron.  While the aggregate properties of all other elements
influences the $\Delta\chi^2$ of the solution, the strengths of their
individual lines are not significant.  These elements and their ions
simply do not have the cross-section or column densities to individually
affect the LIP solution.  In particular, we have no discriminatory power
for helium, aluminum, silicon, argon, calcium, and nickel.  A few
remaining elements supported by PHASE are marginally seen, with slight
evidence for neon absorption (though the data are also consistent with
the continuum at these wavelengths).  Possible magnesium absorption is
also seen, but these lines are at a location in the spectrum where the
continuum is not well determined; as a result, Mg absorption does not
have any diagnostic power.

We find that absorption lines of C, N, and O (\ion{C}{5}, \ion{N}{6},
\ion{O}{5}, \ion{O}{6}, and \ion{O}{7}) are dominant contributers to the
LIP, with tens of $\Delta\chi^2$ each, while iron contributes
substantially with its UTA. On the other hand, the Carbon lines
(\ion{C}{5} and \ion{C}{6}) do not contribute significantly to the HIP.
Oxygen and iron are the primary drivers in the need for an HIP. For this
reason, we allow the metallicity of C, N, O, and Fe to vary but fix all
other abundances to solar. This fit provides an additional
$\Delta\chi^2\sim31$ and shifts the absorber parameters to $\log
U_{LIP}=-0.73$, $\log N_{H,LIP}=19.66$, and $\log U_{HIP}=+0.93$, $\log
N_{H,HIP}=19.50$. The improvement in the fit is again significant
according to the F-test (for four additional degrees of freedom, the
abundances of C, N, O, and Fe, the F-test confidence is 99.999\%). We
refer to this solution as our model 3 (Table 1).

All the four elements require some degree of super-solar abundance in
this solution with abundances from about $2_{\odot}$ for carbon,
$5_{\odot}$ for nitrogen, $7_{\odot}$ for iron to $8_{\odot}$ for
oxygen.  The parameters of these fits are given in Table~\ref{tab:abs}
and abundances are given in Table~2. Even though the spectral fit is
statistically better, this is not a robust result because the hydrogen
column density in the warm absorber is not well constrained through
X-ray observation. The X-ray data, however, do find non-solar abundance
mixture, which is independent of the total overall metallicity. The best
fit finds ``super-solar'' abundances for C,N,O and Fe because abundances
of all other elements were fixed to solar. For this reason, we downplay
this result based on X-ray data alone. However, the stability analysis
and inclusion of FUV data (\S 4) makes the case for super-solar
abundances stronger.


\section{Discussion} \label{sec:conclude}

\subsection{Properties of the warm absorber}

One somewhat surprising result is that we find that neon is
under-abundant with respect to oxygen.  We report only upper limits on
the neon column because the significance of detection for any one line
is small, especially if the continuum is not adequately determined.
Even so, the limits we are able to set are meaningful if only because of
the high abundance of oxygen.  We find that the gas in the circumnuclear
regions of Mrk\,279 has a Ne/O ratio $<0.10$ if not $<0.04$ with respect
to Solar (model 3). This is different from the intergalactic
medium around the Galaxy where Ne is found to be overabundant with
respect to oxygen (\citealt{rikneon}, \citealt{pksneon}).

The total hydrogen column density adding together the two phases
detected in the X-ray is about $10^{20}$\,cm$^{-2}$.  To our knowledge,
this is the smallest column density of all known AGN outflows observed
in X-rays.  It is unlikely that there is a significant amount of column in
any other intermediate temperature phase as all of the major absorption
features are reasonably fit by the LIP and HIP combination (however, we
cannot rule out the presence of a third, hotter, component producing
absorption lines \ion{Fe}{24} to \ion{Fe}{26} as LETG data are not
sensitive to these lines).  In this sense, Mrk 279 contains the weakest
absorber of all, and perhaps because of its weakness, it allows
metallicity diagnostics. If the absorber column densities are large,
there is a heavy blending of lines from different elements or even from
different charge states of the same element, making it difficult to
separate metallicity from total column density. In Mrk 279, however, the
total column density is low, so only the strong lines are present, thus
there is no blending and lines are also not saturated. This led us to
find good evidence for super-solar abundances of C, O, N and Fe.

In NGC~3783, the LIP and HIP components were found to be in pressure
balance. To test whether the same is true in Mrk 279, we generated a
pressure--temperature equilibrium curve (the so called ``S'' curve) for
the assumed spectral energy distribution which is shown in Figure 3. The
solid red curve corresponds to absorbers with solar
metallicity. Interestingly, we find that the LIP and HIP are not at the
same pressure in this case. Another interesting thing about the red
curve is that it shows a monotonic rise in pressure with temperature; it
has no equilibrium zone where components of different temperatures can
exist in pressure balance with each other. Such behavior is seen in
sources with steep soft X-ray spectra (e.g. NGC~4051, Komossa \& Mathur
(2001)). Indeed, in Mrk~279, the continuum fit has a slope of
$\Gamma\approx 2$, which may be the cause of the observed shape of the
equilibrium curve. It should be noted, however, that with LETG data, we
are not sensitive to the harder X-ray range, so it is possible that the
intrinsic continuum of Mrk~279 is flatter, and in that case, the
true ``S'' curve will have an equilibrium zone. Given the observed
spectral shape, however, there is no equilibrium zone and the LIP
and HIP components are not in pressure balance. Komossa \& Mathur (2001)
showed that the shape of the equilibrium curve not only depends upon the
spectral energy distribution of the source, but also on the metallicity
of the absorber and that super-solar abundances restore the equilibrium
zone in steep spectrum sources. Since we have some evidence of
super-solar abundances in Mrk~279, we generated a new
pressure--temperature with super-solar abundances of C, O, N and Fe as
observed; this is shown as the solid blue curve in figure 3. As
expected, the new solution develops an equilibrium zone and, within the
uncertainties, the LIP and HIP components are now at the same
pressure. This exercise makes the case for super-solar abundances
stronger.

\subsection{Comparison with the UV absorber}

We can also compare our results to that of \citet{gabel} who looked at
the absorption in UV wavebands covered by FUSE and HST simultaneously
with the X-ray observations. The UV/FUV resolution is much higher than
X-ray resolution, and each absorption line was found to be made of
multiple kinematic components (labeled 2, 2a and 4a by
\citet{gabel}). These authors report  column densities for ``2+2a''
and ``4a'' components, each in \ion{H}{1}, \ion{C}{4}, \ion{N}{5}, and
\ion{O}{6} for two models (A \& B). In model A, a single covering factor
was assumed to describe all lines while in model B, independent covering
factors for the continuum source and emission lines were assumed. Since
none of the kinematic components can be resolved in the X-ray spectrum
of Mrk\,279, we will consider total column density of each ion in their
two models A \& B.  Since \citet{gabel} do not discuss a photoionization
model that best fits their ionic column densities, and since PHASE does
not yet have the capacity to fit UV/FUV spectra simultaneously with
X-ray spectra, we construct photoionization models with Cloudy
\citep{cloudy} and look for solutions to each of their two models to
test the consistency with our X-ray results.  Without detailed knowledge
of the data reduction process we assign conservative 0.1dex
uncertainties to each of their column density measurements, since the
errors are not reported in \citet{gabel}.

We create Cloudy models with solar abundance and mixtures and with the
Mrk 279 SED (consistent with the PHASE models above) and search for
models with predictions consistent with column densities of all four
ions observed in the UV.  In \citet{gabel}, the \ion{C}{4} and
\ion{H}{1} columns are measurements, while the \ion{N}{5} and \ion{O}{6}
columns are lower limits in most of the models.  We then note which
$\log U$--$\log N_H$ values produce models with column densities
consistent with observations; this process is similar to that presented
in \citet{f05a, f05b}.  Recall that we have assigned uniform
conservative uncertainties to these column densities and thus the
best-fit model should only be considered ``plausible''. Let us first
focus on Model B which is the preferred model of \citet{gabel}. We find
that the HIP solution does not contribute significantly to the UV
absorber. This is no surprise and this result is similar to that found
in other AGNs, e.g. NGC 3783. As in other AGNs, it is possible that the
LIP solution is related to the UV absorber, so we compared the
predictions from the LIP solution of our model 2 (two absorber solution
with solar abundance) with UV column densities as in Model B. To out
surprise we found that the the X-ray model significantly overproduces UV
column densities for either ``$2+2a$'' or ``$2+2a+4a$'' systems (figure
~\ref{fig:b2}).

Next we tried our model 3 (two absorber solution with variable
abundance) for comparison with the UV data. In this case, the LIP solution
has elemental abundances $2_{\odot}$ for carbon, $5_{\odot}$ for nitrogen,
and $8_{\odot}$ for oxygen (\S 3.2). The \ion{C}{4}, \ion{N}{5} and
\ion{O}{6} column densities were adjusted accordingly to reflect the
super-solar abundances in producing the $\log U$--$\log N_H$ plot. The
result is shown in figure~\ref{fig:b4} and it clearly shows that the LIP
solution (marked with an open star) is consistent with the contours of
allowed values of the UV data for ``$2+2a+4a$'' systems (model B). Note
that the \ion{N}{5} and \ion{O}{6} contours are lower limits, so all the
values toward upper left of the contours are allowed. \ion{H}{1} is not
detected in system $4a$, so the allowed range is for system ``$2+2a$''
only.

The above analysis leads to two results: (1) the LIP component of the
X-ray absorber and the UV absorber appear to arise in the same part of
the nuclear outflow of Mrk 279 and (2) there is evidence of super-solar
metals in the outflow. The first result is interesting in that it
unifies the X-ray and UV absorbers (e.g. Mathur et al. 1994, 1995), but
is not particularly surprising given that such a relationship is now
seen in several AGNs. The second result is, however, surprising. The
PHASE analysis of the X-ray spectrum itself pointed toward this
solution, but given the uncertainty in various parameters we did not
consider this as a robust result, even though it was statistically
significant. The unification of the X-ray/UV data, however, {\it
requires} super-solar abundances of metals, making this a much more
robust result.

\subsection{Comparison with Costantini et al. and Arav et al.}

The parameters of the ionized outflow that we find are different from
the preliminary report by \citet{c05} (\S 1). The total column density
of the ionized absorber that we find ($\log N_H=20.0$) is lower by over
two orders of magnitude for both the LIP and HIP. We note that our
result is consistent with the UV value in \citet{gabel}. A total column
density larger than $10^{22}$ cm$^{-2}$, as in \citet{c05}, is clearly
ruled out. In the final stages of our work we found that \citet{c06}
have published their work. Our new result agrees with their new result
in that they too find the column density of the absorber to be about
$\log N_H=20.0$ with two distinct components of low- and high-ionization
parameter, similar to our LIP and HIP. \citet{c06} find somewhat
different outflow velocities for their two components ($-202\pm50$ and
$-500\pm130$ km s$^{-1}$). We find the outflow velocity of $290\pm 50$
km s$^{-1}$, similar to that of their low-velocity component, and the
LIP and HIP have similar velocities within the uncertainties
(2$\sigma$). The \citet{c06} paper does not discuss the relation between
the X-ray and UV absorbers or the evidence for super-solar carbon,
nitrogen and oxygen. They rule the presence of a ``compact'' Mrk~279
outflow in which LIP and HIP are in pressure balance based on analysis
similar to what is shown in figure 3. As we show, however, that the
super-solar abundances alleviate this problem (and, perhaps, a flatter
intrinsic continuum). The present observations are thus fully consistent
with a compact absorber in Mrk 279.

After we submitted our paper, we became aware that Arav et al. (2007)
have independently arrived at conclusions similar to ours in that the
CNO abundances in Mrk~279 are super-solar.\footnote{Their preprint
appeared on the electronic preprint archive astro-ph.} These authors derived
abundances based on UV data and found their result to be consistent with
the X-ray data; in this sense their approach is complementary to ours and
it is good to see that two different methods gave similar results
overall. The exact abundance values are consistent for carbon and
nitrogen (about 2.4 and 4.5 solar respectively), but differ by 3$\sigma$
for oxygen (we get $8.4^{+2.2}_{-2.3}$ while Arav et al. report
$1.6\pm0.8$ solar). Arav et al. method is not sensitive to iron
abundance, which we find to be $7.4^{+5.0}_{-1.7}$ solar. 

\section{Conclusion}

The column density of the outflow in Mrk\,279 
is much lower than that in other AGNs, e.g. NGC\,3783 \citep{phase}, 
NGC\,5548 \citep{m5548}, or NGC\,4051 \citep{k06}. In this sense, the
outflow in Mrk\,279 is the weakest one we know. The column density does
not seem to be correlated to the source luminosity; the luminosity of
NGC\,4051 is a factor of one hundred lower than that of Mrk\,279.

By itself, the evidence of super-solar abundances in the X-ray spectrum
of Mrk~279 is not strong. However, the unification of UV and X-ray
absorbers and the pressure balance between the two components (LIP and
HIP) of the outflow, both require super-solar abundance. This
strengthens the case for it.

Among all the AGNs for which high resolution X-ray spectra are
available, Mrk 279 is the first case in which super-solar elemental
abundances are found . This may be related to the fact that Mrk 279
contains the weakest absorber, so the absorption lines are not
blended. This allows us to separate the effect of metallicity from that
of column density, which was difficult to do in other AGNs with stronger
absorbers. The only other AGN in which a robust metallicity measurement
is made, based on absorption line study, is Mrk 1044
\citep{f05b}. Mrk 1044 is a narrow-line Seyfert 1 galaxy (NLS1) and the
super-solar metallicity in its nuclear outflow was related to its being
a NLS1. The evidence of super-solar metallicity in Mrk 279, a normal,
low luminosity Seyfert galaxy, makes us wonder if high metallicities are
common in circumnuclear regions of AGNs. If this is the case, then
strong outflows may contribute significantly toward enriching the
intergalactic medium.

Since we do not know the exact geometry and location of the warm
absorber in Mrk\,279, it is difficult to calculate its rate of mass and
energy outflow. The location of the warm absorber in NGC 4051 is found
to be 0.5--1.0 light days from the central source (Krongold et
al. 2006). If the distance of the warm absorber from the nucleus scales
as the square root of the luminosity, then, it would be about 5--10
light days for Mrk\,279. Assuming a covering fraction of 10\%, the mass
outflow rate is $\sim10^{-6}$\,M$_{\odot}$ yr$^{-1}$ and the kinematic
luminosity is $\sim3\times10^{34}$\,erg s$^{-1}$.  This wind is a
negligible component of the AGN both bearing away a small fraction of
the total mass accretion rate ($10^{-5}$) and carrying a small fraction
of the radiative luminosity ($4\times10^{-11}$).

To summarize, we find that the warm absorbing outflow in Mrk\,279 has
two components in pressure balance with each other. The X-ray and UV
absorbers appear to be parts of the same overall outflow. Super-solar
metal abundances are required to reconcile the X-ray and UV absorber
properties. The outflow is very weak, however, and is unlikely to make
any significant impact on the energy balance in the host galaxy or the
surrounding intergalactic medium.

\begin{acknowledgements}

\end{acknowledgements}

\clearpage

\begin{deluxetable}{lccccccc}
\tabletypesize{\footnotesize}
\tablecaption{X-ray Absorber Parameters\tablenotemark{a}
\label{tab:abs}}
\tablehead{
\colhead{Absorber\tablenotemark{b}}
&\colhead{$\log U$}
&\colhead{$\log N_H$}
&\colhead{$\log N_C$}
&\colhead{$\log N_N$}
&\colhead{$\log N_O$}
&\colhead{$\log N_{Fe}$}
&\colhead{$\chi^2$/d.o.f.}}
\startdata
Model 1& $-0.36$ & 19.94 & 16.49 & 15.91 & 16.81 & 15.45 & 2888/3195 \\
&&&&&&& \\
Model 2: LIP& $-0.95$ & 20.19 & 16.74 & 16.16 & 17.03 & 15.70 & 2845/3192 \\
Model 2: HIP& $+0.45$ & 19.5 & 16.27 & 16.05 & 17.26 & 15.72 &... \\
&&&&&&& \\
Model 3: LIP& $-0.73$ & 19.66 & 16.61 & 16.29 & 17.46 & 16.04 & 2814/3188 \\
Model 3: HIP& $+0.93$ & 19.5 & 16.44 & 16.13 & 17.29 & 15.88 &... \\
\enddata
\tablenotetext{a}{ Column densities are in $\log$\,cm$^{-2}$. A Mrk 279 SED was used in all calculations}
\tablenotetext{b}{ Model 1 $\equiv$ Single absorber, Solar metallicity \\
~ ~ Model 2 $\equiv$ Two absorbers, Solar metallicity \\
~ ~ Model 3 $\equiv$ Two absorbers, CNOFe variable abundances.}
\end{deluxetable}

\begin{deluxetable}{lc}
\tabletypesize{\footnotesize}
\tablecaption{Best fit abundances\tablenotemark{a}
\label{}}
\tablehead{
\colhead{Element}
&\colhead{Abundance relative to Solar}
}
\startdata
Carbon & 2.5$^{+0.9}_{-0.5}$ \\
Nitrogen & 4.6$^{+2.4}_{-1.4}$ \\
Oxygen & 8.4$^{+2.2}_{-2.3}$ \\
Iron & 7.4$^{+5.0}_{-1.7}$ \\
\enddata
\tablenotetext{a}{For Model 3}
\end{deluxetable}

\begin{figure}
\epsscale{1.0}
\plotone{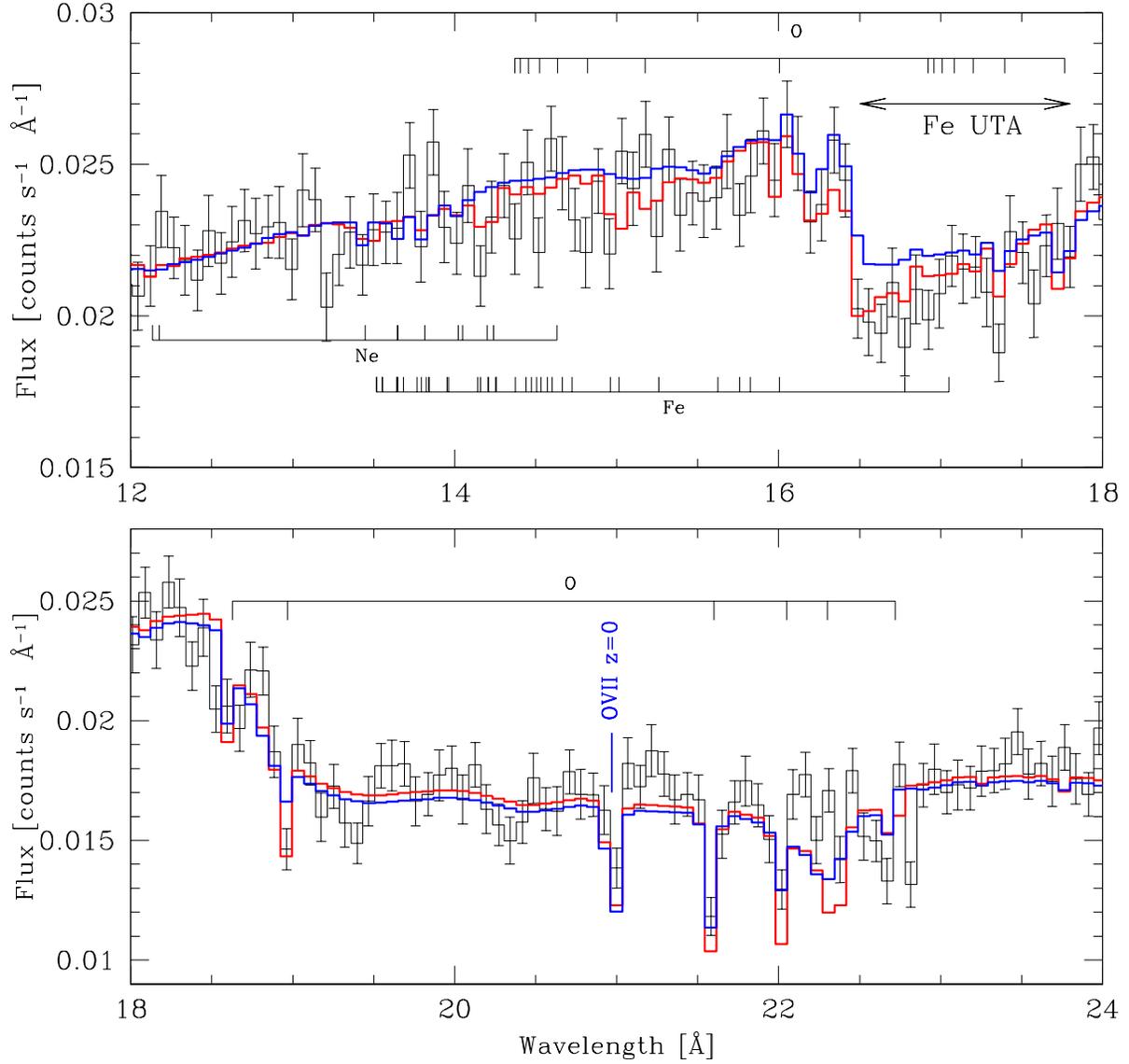}
\caption{\label{fig:spectrum}
The LIP (blue curve) and LIP$+$HIP (red curve) models (models 1 \& 2
respectively) fitted to the LETG data of Mrk~279 (black histogram) in
the spectral range of interest in the Mrk~279 rest frame. Lines from the
z=0 absorption system were also added to the model.  }
\end{figure}

\begin{figure}
\epsscale{1.0}
\plotone{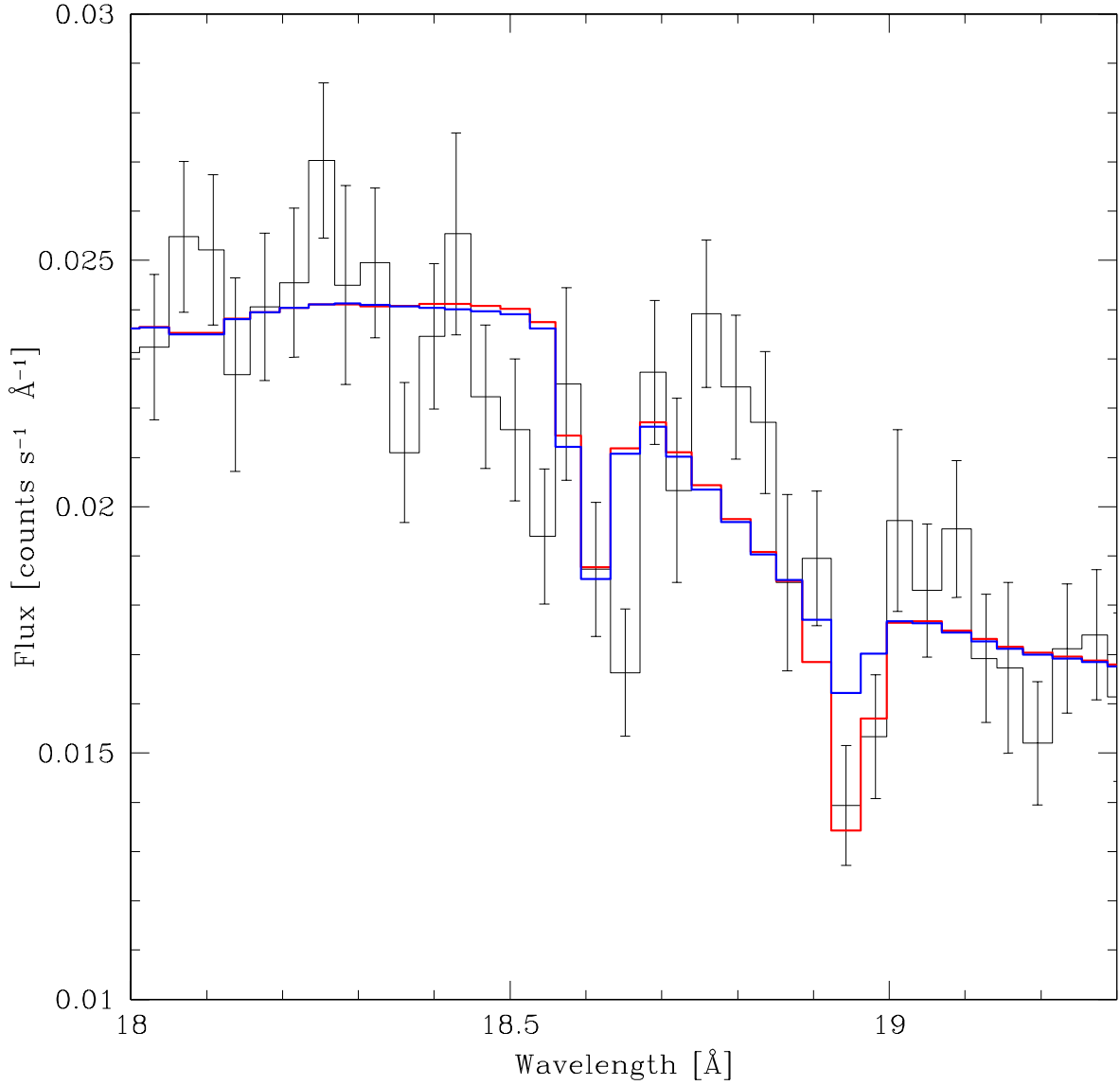}
\caption{\label{fig:oviii}
Zoomed in view of the \ion{O}{8} line at 18.969\AA\ (rest frame) showing that
model 1 (blue curve) under-predicts the line strength and that model 2
(red curve) is a significantly better fit. 
}\end{figure}

\begin{figure}
\epsscale{1.0}
\plotone{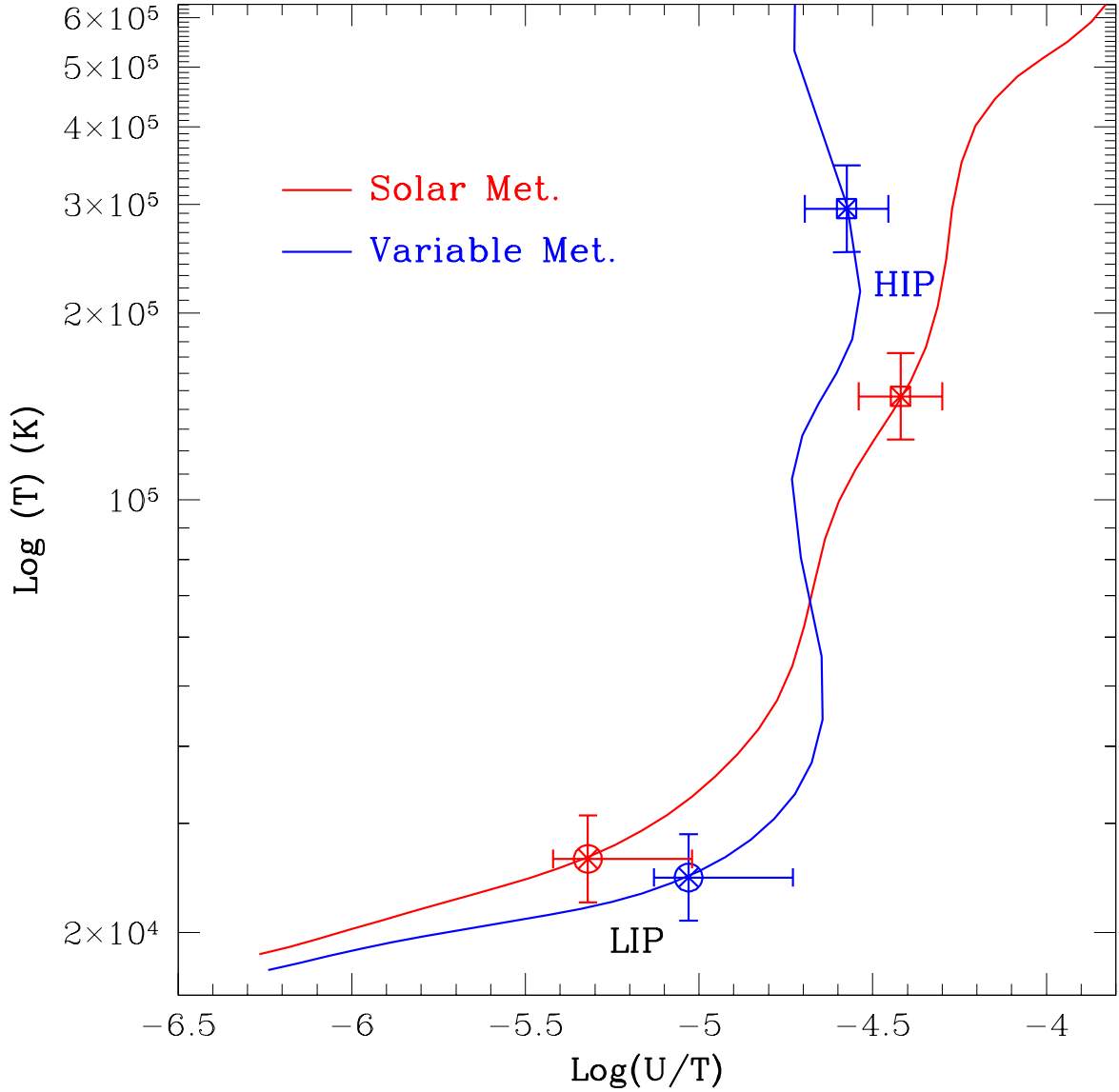}
\caption{\label{fig:scurve}
The pressure--temperature plot for the Mrk~279 SED. The red curve is for
solar abundances while the blue curve is for super-solar C,N,O as
observed. The LIP and HIP components (shown as circles with a cross on
each curve) do not correspond to the same pressure (plotted as $\log(U/T)$)
on the red curve but are consistent with the same pressure on the blue
curve. Note also that there is no equilibrium zone in the red curve, but
there is one in the blue curve. 
}\end{figure}

\begin{figure}
\epsscale{0.7}
\plotone{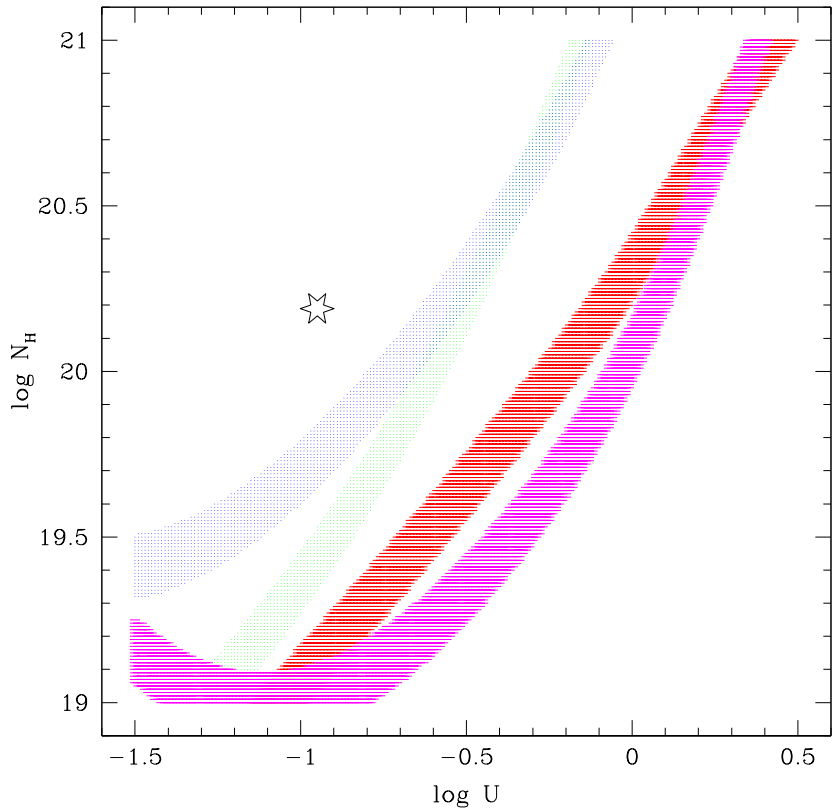}
\caption{\label{fig:b2}
A $\log U$--$\log N_H$ plot generated using Cloudy models for a Mrk 279
input spectrum and solar metallicity. The contours correspond to
observed ionic column densities reported by \citet{gabel}, their Model
B, systems ``$2+2a$'' plus ``$4a$''. Here Red = \ion{H}{1}, Green =
\ion{C}{4}, Blue = \ion{N}{5}, Purple = \ion{O}{6}.  The \ion{H}{1} and
\ion{C}{4} columns are for measurements while \ion{N}{5} and \ion{O}{6}
are lower limits for which the allowed parameter space is to the upper
left side of the corresponding contour. The open star represents the LIP
solution in Model 2. While the \ion{N}{5} and \ion{O}{6} lower limits
are consistent with the LIP solution, \ion{C}{4} and \ion{H}{1} are
clearly over-predicted. The LIP of Model 2 thus does not correspond to the UV
absorber. 
}\end{figure}

\begin{figure}
\epsscale{0.7}
\plotone{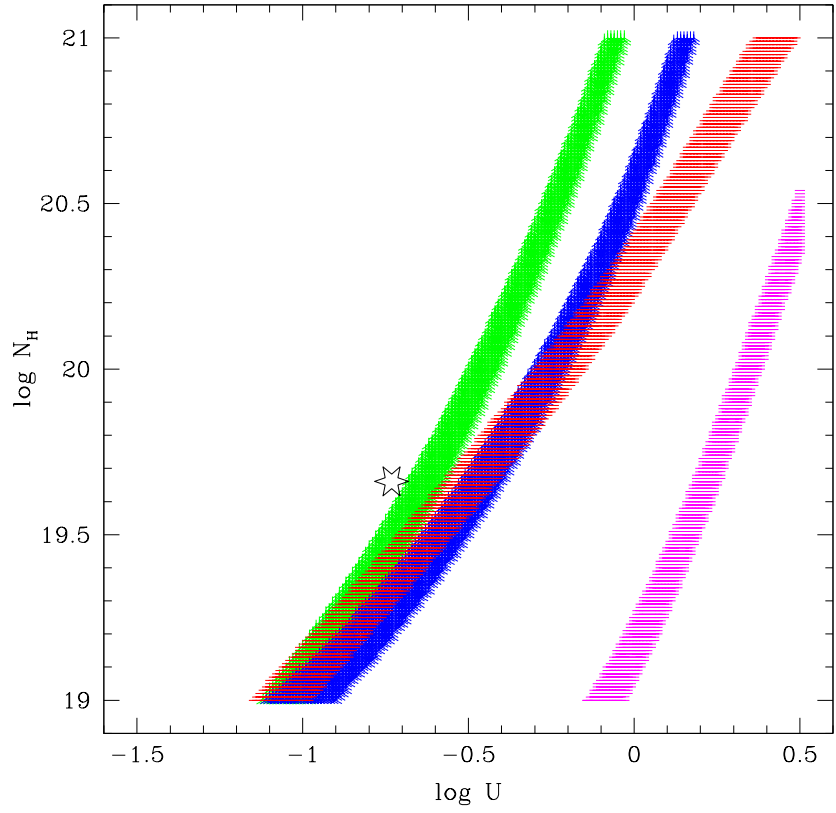}
\caption{\label{fig:b4}
Same as in figure~\ref{fig:b2}, but for the LIP of Model 3. In this
model, C, N, and O abundances are found to be 2, 5, and 8 times solar
respectively. Now we see that the X-ray solution (open star) is
 consistent with the UV data.
}\end{figure}

\end{document}